# Annealing-induced long-range charge density wave order in magnetic kagome FeGe: fluctuations and disordered structure

Chenfei Shi [1] [††], Yi Liu [2, 3] [††], Bishal Baran Maity [4], Qi Wang [5, 6], Surya Rohith Kotla [7], Sitaram Ramakrishnan [8], Claudio Eisele [7], Harshit Agarwal [7], Leila Noohinejad [9], Qian Tao [3], Baojuan Kang [1], Zhefeng Lou [10], Xiaohui Yang [11], Yanpeng Qi [5, 6, 12], Xiao Lin [10], Zhu-An Xu [3, 13], Arumugam Thamizhavel [4], Guang-Han Cao [3, 13], Sander van Smaalen [7] [*], Shixun Cao [1, 14] [*], and Jin-Ke Bao [1, 14, 15] [*]

[1] *Department of Physics, Materials Genome Institute, Institute for Quantum Science and Technology, Shanghai University, Shanghai 200444, China*
[2]*Department of Applied Physics, Zhejiang University of Technology, Hangzhou 310023, China*
[3]*School of Physics, Zhejiang Province Key Laboratory of Quantum Technology and Devices, Zhejiang University, Hangzhou 310027, China*
[4]*Department of Condensed Matter Physics and Materials Science, Tata Institute of Fundamental Research, Mumbai 400005, India*
[5]*School of Physical Science and Technology, ShanghaiTech University, Shanghai 201210, China*
[6]*ShanghaiTech Laboratory for Topological Physics, ShanghaiTech University, Shanghai 201210, China*
[7]*Laboratory of Crystallography, University of Bayreuth, Bayreuth 95447, Germany*
[8]*Department of Quantum Matter, AdSE, Hiroshima University, Higashi-Hiroshima 739-8530, Japan*
[9]*P24, PETRA III, Deutsches Elektronen-Synchrotron DESY, Hamburg 22607, Germany*
[10]*Key Laboratory for Quantum Materials of Zhejiang Province, School of Science, Westlake University, Hangzhou 310024, China*
[11]*Department of Physics, China Jiliang University, Hangzhou 310018, China*
[12]*Shanghai Key Laboratory of High-resolution Electron Microscopy, ShanghaiTech University, Shanghai 201210, China*
[13]*Collaborative Innovation Centre of Advanced Microstructures, Nanjing University, Nanjing 210093, China*
[14]*Shanghai Key Laboratory of High Temperature Superconductors, Shanghai University, Shanghai 200444, China*
[15]*School of Physics, Hangzhou Normal University, Hangzhou 311121, China*

[*]Corresponding author (Sander van Smaalen, email: smash@uni-bayreuth.de; Shixun Cao, email: sxcao@shu.edu.cn; Jin-Ke Bao, email: baojk7139@gmail.com)

[††]These authors contributed equally to this work.

**Abstract**: Charge density wave (CDW) in kagome materials with the geometric frustration is able to carry unconventional characteristics. Recently, a CDW has been observed below the antiferromagnetic order in kagome FeGe, in which magnetism and CDW are intertwined to form an emergent quantum ground state. However, the CDW is only short-ranged and the structural modulation originating from it has yet to be determined experimentally. Here we realize a long-range CDW order by post-annealing process, and resolve the structure model through single crystal x-ray diffraction. Occupational disorder of Ge resulting from short-range CDW correlations above $T_{CDW}$ is identified from structure refinements. The partial dimerization of Ge along the $c$ axis is unveiled to be the dominant distortion for the CDW. Occupational disorder of Ge is also proved to exist in the CDW phase due to the random selection of partially dimerized Ge sites. Our work provides useful insights for understanding the unconventional nature of the CDW in FeGe.

**Introduction**

Charge density wave (CDW) is a collective phenomenon of electrons which form charge density modulation usually accompanied with periodic lattice distortion [1]. It is commonly observed in quasi-one-dimensional materials due to nearly perfect Fermi-surface nesting, known as the Peierls instability [2]. Quasi-two-dimensional and even three-dimensional materials can also exhibit CDW order under various mechanisms, for instance, partial Fermi surface nesting or q-dependent electron-phonon coupling [3]. In response to CDW orders, the atoms in the original structure undergo displacements to form dimers, trimers or other clusters [4]. In addition, electronic transport properties can change dramatically during the CDW transition because of the gap opening in the electronic band structure [1]. For example, a CDW triggers a metal-insulator transition with thermal hysteresis in 1T-TaS$_2$, providing the potential application in electronic devices [5]. CDWs can also compete or intertwine with other collective phenomena such as superconductivity and spin density waves [6-8]. Thus, it has garnered a great interest in the fundamental research as well as the application of electronic devices.

Materials with a kagome lattice, a two-dimensional network of corner-shared triangles, provide a fertile platform for realizing abundant exotic states such as quantum spin liquid [9] and fractional quantum Hall effect [10]. The kagome lattice is geometrically frustrated, and it supports flat bands, Dirac points and van Hove singularities (vHSs) in electronic structures [11]. A non-trivial CDW has been manifested in the nonmagnetic kagome family $A$V$_3$Sb$_5$ ($A$ = K, Rb and Cs) which is unveiled to exhibit pos-sibly time-reversal symmetry breaking CDW order ($T_{\text{CDW}} \approx 78 - 103$ K) and unconventional superconductivity ($T_c \approx 0.92 - 2.5$ K) [12].

Recently, a CDW order ($T_{\text{CDW}} \approx 100$ K) was discovered in B35-type FeGe with a Fe-kagome lattice below the A-type antiferromagnetic (AFM) order ($T_N \approx 410$ K) [13-16]. The magnetic moment of Fe is enhanced by around 0.05 $\mu_B$ after it enters the CDW state, indicating a strong coupling between CDW and magnetism [13]. The appearance of edge states and anomalous Hall effect below $T_{\text{CDW}}$ suggests that band structure topology is also involved in FeGe [13, 17]. Additionally, topological band-inversion-induced edge mode is experimentally detected in single-layer FeGe nanoflakes on a substrate [18]. Thus, FeGe turns out to be a fascinating system where charge order, topology and magnetism entangle together.

However, the reported CDW in FeGe possesses only a short-ranged order with small coherence length and is also fragile [13, 17, 19, 20]. Can a long-range ordered CDW be realized in a high-quality sample by optimizing the synthetic process? Although several theoretical calculations and experimental studies on the structure of CDW have been carried out [21-25], the consensus of them is far from being reached. What is the explicit modulated structure for the CDW in FeGe?

Motived by these questions, we performed an annealing study on the behavior of the CDW in FeGe and achieved a long-range ordered CDW by post-annealing under certain temperatures. We also resolve the crystal structure of the CDW ordered state in FeGe by single crystal x-ray diffraction (SXRD) with a synchrotron radiation source. The crystals annealed at 573 K exhibit a sharp magnetic-susceptibility drop and a specific-heat jump at $T_{\text{CDW}}$, as well as intense superlattice peaks in SXRD at 80 K below $T_{\text{CDW}}$. Above $T_{\text{CDW}}$, the existence of short-range CDW correlations is inferred from the occupational disorder in the refined crystal structure and diffuse x-ray scattering signals. The dimerization of Ge atoms in the kagome plane is the major distortion for CDW, concomitant with small structural distortions of Fe kagome and Ge honeycomb planes. Because each Ge atom in the kagome plane may dimerize with one of its equivalent neighbors, leading to multiple domains and thus structural disorder in the refined model for the CDW phase.

## 2 Materials and method

**Sample synthesis**

B35-type FeGe single crystals were grown by chemical vapor transport method using iodine as the transport agent. The crystals used in this work came from three different groups under slightly different growth conditions and have varied physical properties according to the magnetic susceptibility and single crystal x-ray diffraction measurements. Crystal label information is provided in Supplementary Materials (SM).

The crystal batches FeGe#A (as-grown) and FeGe#A_$n$ ($n$ = 1-6, annealed) were grown in the synthesis lab of Zhejiang University. The growth was basically reproduced in the lab of Shanghai University, with crystals showing almost the same physical properties. Whether a polycrystalline binary precursor FeGe is used or not, does not play an important role in getting the B35-type FeGe crystals. We found that using the elements as the ingredients leads to bigger crystals as shown in Figure S1. Firstly, iron powder (99.9%) and germanium pieces (99.9999%) were mixed and ground in a stoichiometric ratio, then sealed in an evacuated quartz tube with additional iodine (5 mg/cm$^3$). Secondly, the sealed quartz tube was placed in a two-zone furnace with a temperature gradient from 873 K (educt) to 823 K (product). After 12 days of growth, the tube was removed from the furnace and quenched in cold water. B35-type FeGe single crystals with the size of 1.5 × 1.5 × 2.5 mm$^3$ have been obtained. Finally, the obtained samples were annealed at different temperatures for 10 days.

The crystal batch FeGe#B was grown at the Tata Institute of Fundamental Research, Mumbai. Iron and germanium were taken in a stoichiometric ratio and arc-melted into an ingot in a tetra arc furnace. The ingot was then annealed at 973 K for 7 days after which it was quenched in liquid nitrogen. After being annealed, the ingot was powdered and sealed in a long quartz tube with iodine. The sealed quartz tubes were then loaded into a multi-zone furnace. The hot zone was maintained at 843 K for around 12 days while the cold zone was also maintained at 843 K for around 4 days, then cooled to 793 K at the rate of 5 K/h and finally maintained at 793 K for around 8 days. The whole quartz tube was cooled down to room temperature at a rate of 50 K/h after the crystal growth was finished.

The crystal batch FeGe#C was grown at the ShanghaiTech University, Shanghai. The polycrystalline FeGe precursor was synthesized from a mixture with a molar ratio of Fe: Ge = 1: 1, which was sealed in an evacuated quartz tube, heated up to 1273 K, cooled down to 973 K and then held for 1 week. The FeGe powder ground from the synthesized precursor was sealed with iodine in an evacuated quartz tube. The two ends of the quartz tube were heated to 843 and 818 K, respectively, and then maintained for

3 weeks. The final product was cooled down to room temperature naturally after the power of the furnace was shut off.

**Structure characterization**

Single crystal x-ray diffraction (SXRD) of the as-grown sample FeGe#A was performed on Bruker D8 VENTURE instrument with Mo $K\alpha$ radiation ($\lambda$ = 0.71073 Å). SXRDs of the annealed sample #A_2, and as-grown samples #B and #C with a synchrotron radiation source were carried out at the station of EH2 of beamline P24 of PETRA-III at DESY in Germany. The x-ray wavelength was tuned to 0.5 Å. The temperature of the samples was controlled by open flow helium cryostat. The measurement strategy and data process are similar to the one adopted in reference [26]. The samples were continuously rotated along the phi axis at a speed of 1 degree/second. The Pilatus detector CdTe 1M was used to receive the diffracted photons with a readout frequency of 10 Hz. 3640 frames were recorded for a single complete run with a step size of 0.1 degree and an exposure time of 0.1 s for one frame. 10 repeated complete runs were binned and added for one single complete data set with a step size of 1 degree and an effective exposure time of 10 seconds per frame for the follow-up structural analysis. Experiments with an attenuator were also carried out to avoid the overexposure of the main reflections. The software package Eval15 [27] and CrysAlispro (CrysAlis Pro Version 171.40.67a, Rigaku Oxford Diffraction) were used to reduce the collected data and create the reconstructed images in the reciprocal space, respectively. The program SADABS [28] was used to scale the reflections containing two processed runs with and without an attenuator, and do the absorption correction. The crystal structures were solved by a Superflip method and refined against structure factor $F$ in Jana2006 [29, 30]. The total occupancy of neighboring disordered Ge sites has been constrained to 1 and their atomic displacement parameters are also constrained to be identical for all the final refined models. Crystal structures were plotted using the software VESTA [31]. A desktop scanning electron microscope & energy-dispersive x-ray spectroscopy (EDS) (Hitachi, FlexSEM1000II) was used to check the composition of FeGe crystals (Figure S1 in SM).

**Second harmonic generation measurements**

Second harmonic generation (SHG) experiments were conducted using a confocal microscope (WITec, Alpha300RAS) equipped with a 1064-nm laser excitation source

(NPI Rainbow 1064 OEM). For low-temperature measurements, a cryostat for continuous-flow optical microscopy was employed to detect whether the inversion symmetry is broken upon cooling that leads to a nonzero second-order nonlinear signal. The measurements were performed by directing the excitation beam both along and perpendicular to the *c*-axis (Figure S8 in SM).

**Physical property measurements**

Temperature-dependent magnetic susceptibility was measured using a commercial superconducting quantum interference device (MPMS3, Quantum Design). The samples were mounted on a quartz sample holder with magnetic fields applied along certain crystallographic direction. Resistivity measurements were performed using a commercial physical properties measurement system (PPMS-14 T, Quantum Design DynaCool). The heat capacity measurements were characterized on a commercial physical properties measurement system (PPMS-9 T, Quantum Design) using a thermal relaxation method.

## 3 Results

B35-type FeGe crystallizes in the CoSn-type structure with space group *P*6/*mmm* [32]. The structure of FeGe can be described as the alternating stacking of Ge and $Fe_3$Ge layers along the *c* axis (Figure 1(a)). Each $Fe_3$Ge layer consists of a kagome lattice of Fe atoms with additional Ge atoms (labeled as Ge1) interwoven in the hexagonal voids of the kagome lattice. As for the Ge layer, it forms a honeycomb lattice with the Ge atoms (labeled as Ge2) located directly above or below the center of each $Fe_3$ triangle.

The temperature-dependent magnetic susceptibility of the as-grown crystal of FeGe#A with $H \perp c$ only shows a broad hump at around 96 K (Figure 1(b)), indicating a possible CDW transition, in contrast to a drop in reference [13]. A peak at around 26 K is due to the in-plane component change of the double-cone antiferromagnetic structure below 60 K [15]. The peak temperature varies from one as-grown sample to another under the same growth condition. The as-grown crystals #B and #C using different temperature profiles show a small drop or an anomaly in the magnetic susceptibility at around 100 K but their transitions are broad (Figure 1(b)), indicating the absence of long-range order of the CDW in these as-grown samples, too.

Annealing can be used as an effective way to reduce the concentration of defects in the crystals and promote long-range CDW order like in $CuV_2S_4$ [33]. Thus, we performed the annealing effects study on the CDW order in FeGe. The CDW transition indeed becomes clear for the crystal #A_2 annealed under 573 K, as manifested by a sharp magnetic susceptibility drop at around 112 K. The values of $T_{CDW}$ can fluctuate from 108 to 112 K for crystals from different batches under the same growth condition. They are much enhanced compared with the reported transition temperature (~100 K) in reference [13]. In addition to the pronounced CDW transition in the annealed sample, two anomalies appear in the magnetic susceptibility, at around $T_1$ = 155 and $T_2$ = 262 K, respectively (Figure 1(b)). They are also detected for the as-grown samples #B and #C but absent for the as-grown sample #A. The anomaly at $T_2$ is probably due to the presence of minor cubic B20-type FeGe byproduct, formed either during the annealing or relatively slow cooling process, based on the two following factors: cubic FeGe is expected to be transformed slowly from the hexagonal one under a low-temperature annealing condition according to the phase diagram [34, 35]; the cubic FeGe phase has a helimagnetic transition at around 275-280 K [36, 37] (Figure S2(d) in SM), which roughly matches the value of $T_2$ here. We tried single and powder XRD measurements to directly prove such a hypothesis. Unfortunately, we don't observe extra reflections which can be uniquely attributed to the B20-type phase. The annealing-induced cubic phase might have a form of small clusters as also mentioned in reference [38], which is extremely difficult to be detected by XRD. The as-grown sample #A quenched from the high temperature exhibits no such anomalies in magnetic susceptibility because it avoids the formation of the cubic phase during the cooling process.

We also tried annealing the as-grown samples under different temperatures from 523 to 673 K (Figure 1(c)). The annealing temperature 523 K is not high enough to realize a sharp CDW transition and the annealing temperatures higher than 623 K can significantly reduce the signals from these two anomalies ($T_1$ and $T_2$). We cannot fully eliminate them for all our samples (Figure S2(c) in SM). The two anomalies at $T_1$ and $T_2$ in the magnetic susceptibility can be smeared out by applying a high enough magnetic field (Figure 1(d)), consistent with a net ferromagnetic component in the cubic FeGe (Figure S2(d) in SM). Since $T_1$ is always concurrent with $T_2$ in all our measured samples (Figure 1(b)-(c)) and Figure S2(a)-(c)) in SM), the anomaly at $T_1$ is probably linked to the domain structure of the cubic FeGe. The transition temperature for the double-cone AFM below $T_{CDW}$ also varies from 58 to 68 K for the annealed crystals (Figure 1(b)-(c))).

For both as-grown #A and annealed #A_4 samples, no obvious anomalies at the CDW transition in the resistivity are detected (Figure 2(a)), consistent with previous reports [13, 39]. However, there is a tiny peak at around 109 K (the corresponding transition temperature in magnetic susceptibility) in the derivative of the resistivity for the annealed sample while the anomaly is still absent for the as-grown sample (see the inset in Figure 2(a)), proving that the CDW transition is more pronounced in the former case. We chose the sample annealed at 643 K for a comparison study because it contains the least amount of impurity and probably exhibits the intrinsic behavior. The residual resistivity ratio ($RRR = \rho_{300 K}/\rho_{2 K}$) is estimated to be as low as 7.4 for the as-grown sample and increases to 24.1 for the annealed sample, indicating a smaller concentration of lattice defects in the annealed sample as also evidenced by scanning tunneling microscopy [39]. The long-range CDW order in the annealed sample is well supported by the extremely sharp and large specific heat jump at around 112 K while almost no detectable anomaly at around 96 K is observed for the as-grown sample (Figure 2(b)). The tiny thermal hysteresis and divergent specific-heat jump during the transition (see the inset in Figure 2(b)) suggest a weak first-order transition for CDW [23]. The specific-heat jump and entropy change are obtained as $\Delta C$ = 33.7 K J$^{-1}$ mol$^{-1}$ and $\Delta S_{CDW}$ = 0.76 J K$^{-1}$ mol$^{-1}$, respectively. We do not observe any detectable anomalies in the vicinity of $T_1$ or $T_2$ in the specific heat data, further corroborating that these anomalies come from a minor byproduct of the cubic FeGe.

In order to further understand the annealing effects on CDW within different samples, single crystal x-ray diffraction with a synchrotron radiation source were carried out to capture weak satellite peaks originating from the formation of a supercell [13, 17]. Superlattice peaks are hardly observed at 150 (Figure 3(a) and (b)) and 270 K (Figure S4 in SM) above $T_{CDW}$ for the annealed sample #A_2. They emerge at 80 K below $T_{CDW}$, forming a 2 × 2 supercell in the *ab* plane, see the (*hk*1) plane in Figure 3(c). There are additional superlattice peaks with the indices of (*h* 0 *l*+1/2), indicating a doubled unit cell along *c* axis (Figure 3(d)). All the superlattice peaks at 80 K can be indexed by three independent **q**-vectors: (0.5, 0, 0), (–0.5, 0.5, 0) and (0, 0, 0.5), consistent with previous reports [13, 23]. No observable superlattice reflections are identified below $T_{CDW}$ in the as-grown samples #B (Figure 3(e)) and #A (Figure S3 in SM) while superlattice reflections from a 2 × 2 × 2 supercell are also observed in the as-grown sample #C (Figure 3(f)). However, the intensity of them in the as-grown sample #C is much weaker than the corresponding ones in the annealed sample #A_2 and is not observable for (0 0 $l$+1/2) (Figure S7 in SM), confirming short-range CDW order in the former sample.

In addition, the major diffuse scattering features near to the Bragg reflections observed at 150 K persist down below $T_{CDW}$ even when a long-range CDW order is formed (Figure 3(a), (c), (g) and h as well as Figure S4 in SM), in contrast to most of the diffuse scattering above $T_{CDW}$ growing into sharp superlattice peaks below $T_{CDW}$ in other kagome systems such as Cs(Rb)V$_3$Sb$_5$ and ScV$_6$Sn$_6$ [40-42]. This indicates that extrinsic correlated disorder exists as local defects in the sample as shown below by the refined structure [43]. Some diffuse scattering also transforms into sharp superlattice peaks such as (–2 3.5 2) at 80 K (Figure 3(i) and (j)), pointing to the existence of CDW fluctuations above $T_{CDW}$ in FeGe.

A structure refinement was done on these samples at temperatures above $T_{CDW}$ in order to further understand the contrasting behaviors in samples grown under different conditions. As for the annealed sample #A_2 at 150 K, the initial refinement using the originally reported crystal structure [44] gives an occupancy of 0.95 on the Ge1 site, consistent with reference [13]. However, a significant residual electron density close to the Ge1 site along the $c$ axis exists in the difference Fourier map. Introducing extra Ge atom to the sites (see Figure 4(a) and 4(b)) significantly reduces the figure of merit ($R_{obs}$) from 2.23% to 0.78%. The total occupancy of these three neighboring Ge atom sites (Ge1_1: 0.94, 2×Ge1_2: 0.06) is almost equal to 1 without enforcing a constraint. It is a reasonable result since these three sites are exclusive to each other due to the extremely small interatomic distances (~ 0.7 Å). All the samples investigated in this work with or without the annealing process exhibit a generic behavior with disorder at the Ge1 site (Table S1-S12 in SM). The annealed sample #A_2 has the lowest $R_{obs}$ compared with other as-grown samples, indicating less defects within the crystals due to the annealing process as also vividly visualized by STM tomography in another work [39]. The actual position and occupancy of the additional Ge sites (Ge1_2) slightly vary with different samples. As for the CDW phase, we tried different space groups as provided in references [21, 22] to refine the data in a 2 × 2 × 2 supercell. Disorder inferred by the diffuse scattering at 80 K is introduced in the model (Figure 3(c)). Although different space groups give a marginal difference in $R_{obs}$ (around 2.1-2.5%) for all the reflections, $P6/mmm$ gives a better fit than other options ($P–62m$, $P6mm$, $P–6m2$ and $P622$) in terms of the fit for the superlattice reflections ($R_{obs}$ = 3.47%) and no additional warning information (Table S7 in SM). Moreover, no optical second harmonic generation signal originating from the breaking of inversion symmetry was detected below $T_{CDW}$ (Figure S8 in SM), further corroborating the centrosymmetric space group $P6/mmm$ instead of other non-centrosymmetric ones in FeGe.

The structural model and distortions for the CDW are illustrated in Figure 4(c)-(e). The crystal data and structure refinement are given in Table 1. Ge1_1 site in the original unit cell results in two independent sites (Ge1_1_1 and Ge1_1_2) in the supercell. They have different occupancies of 0.89 and 0.39 (Table S8 in SM), respectively. The independent site Ge1_2_1 in corresponding to the weak disorder from Ge1_2 site in the high temperature increases its occupancy to 0.53. This leads to a significant portion of dimerization of Ge1_2_1 (0.68 Å in deviation from the Ge1_1_2 site) with an interatomic distance of 2.67 Å in the crystal structure. Fe atoms in the kagome lattice have two independent crystallographic sites Fe1_1 and Fe1_2. The distortion of Fe kagome lattice is mainly along the $c$ axis (0.013 Å) with tiny in-plane (less than 0.005 Å) displacement (Figure 4(d)), much smaller than the case of Ge1 dimerization. This is in stark contrast to the dominant structural distortion of V kagome planes for CDW in $A$V$_3$Sb$_5$ [45] and thus suggests that Fermi-surface nesting between the saddle points of the VHSs is not the major instability to drive the CDW order [46, 47]. The two Fe kagome sublattices in the unit cell are symmetry-related by a mirror plane at $z$ = 0.5. There are three different Fe-Fe bond distances (2.488, 2.493 and 2.497 Å) in the modulated structure. The two neighboring honeycomb Ge planes give an opposite displacement direction forming a Kekulé pattern with small in-plane displacements (~ 0.044 Å) only (Figure 4(e)) [48]. Four distinct Ge-Ge distances (2.8477 and 2.9063 Å, 2.8329 and 2.9210 Å) in two different Ge honeycomb planes exist in the modulated structure. The structure model obtained here basically matches the theoretical predictions in reference [21], where CDW order mainly comes from the dimerization of Ge1 sites for magnetic exchange energy saving and leads to concomitant geometric distortions of other sublattices. The random selection of the dimerization of Ge1 sites and their arbitrary displacement configurations from the triple-well energy profile [24] inevitably produce structural disorder in a measured sample when local defects exist as domain walls. The average occupancy of Ge1 sites without undergoing the dimerization at 80 K decreases from 0.95 at 270 K to around 0.76 (3/4o(Ge1_1_1)+1/4o(Ge1_1_2)) which is close to the ideal value of 0.75 according to the dimerization model in a 2 × 2 × 2 supercell [21], for the sample #A_2, suggesting a bulk nature of CDW. The corresponding values obtained from the refinement of the average structure for the as-grown samples #B and #C are around 0.83 below $T_{CDW}$, indicating a large portion of sample domains without entering a 2 × 2 × 2 CDW order. This analysis is also consistent with the results of magnetic susceptibilities and superlattice reflections where CDW signal is less pronounced in the samples #B and #C. As a result, the average occupancy of Ge1 sites without undergoing the dimerization can be used as a criterion to estimate the CDW volume in the sample.

## 4 Discussion and conclusions

The occupational disorder of Ge1 sites in the refined crystal structure above $T_{CDW}$ refers to the existence of residual dimerization of Ge1 sites, which is observed in all the measured samples due to the energy saving from the Ge1 dimerization in the magnetic structure as long as the sample exhibits an A-type AFM [21]. This demonstrates robust short-range CDW correlations above $T_{CDW}$ in all the FeGe samples. Long-range CDW order can precipitate from a number of small incoherent CDW domains during cooling when the concentration of defects is significantly reduced for an annealed sample, suggesting an order-disorder-type CDW transition in FeGe [40, 49]. Multiple CDW domains with different distortion modes coexist at 80 K below $T_{CDW}$ because local defects in the sample can act as the breaker of coherence in forming a single domain structure [50], resulting in a disordered structure in the refinement. However, only short-range CDW order with small domain sizes forms for the as-grown samples when a significant amount of defects act as the domain walls to prevent true coherent long-range CDW [13, 19].

In summary, we have optimized the synthesis and successfully achieved long-range CDW order in B35-type annealed FeGe crystals. Short-range CDW correlations are also evidenced by the observed diffuse x-ray scattering and refined disordered structure model above $T_{CDW}$. Resolved structural model in the CDW phase supports the scenario of the partial Ge1 dimerization along the *c* axis for the formation of CDW but with disorder on the Ge1 sites due to the different modes of the dimerization process.

Our work will not only provide a detailed synthesis recipe to obtain high-quality samples of FeGe for future experiments but also open a route to the landscape of CDW fluctuations from the intertwining of magnetism in this system.

Note added: during the preparation of this work, we became aware of related work by Ziyuan Chen *et al.* (arXiv: 2307.07990) [39] and Xueliang Wu *et al.* (arXiv: 2308.01291) [38] reporting the similar annealing effect on achieving the long-range CDW order in FeGe. All the three independent works demonstrate that Ge dimerization along the *c* axis is the dominant structural distortion for the CDW phase by doing a structural refinement. Reference [39] mainly focuses on the microscopic differences between as-grown and annealed samples by using the STM technique. Reference [38] performs a detailed annealing study on the CDW order under varying temperatures and durations, and unveils that the CDW order is annealing-tunable and closely related to its structural difference on the disorder site of Ge1. Our work conducts a thorough analysis on the structures of the CDW and high-temperature phase for both as-grown and annealed samples by using synchrotron-source XRD. The high-resolution data allow us to determine such subtle changes in the distorted structure with a much higher accuracy for the CDW order. Furthermore, CDW fluctuations above $T_{\text{CDW}}$ are proved by diffuse x-ray scattering and structural refinement in our work.

## 5  Figures and tables

### 5.1  Figures

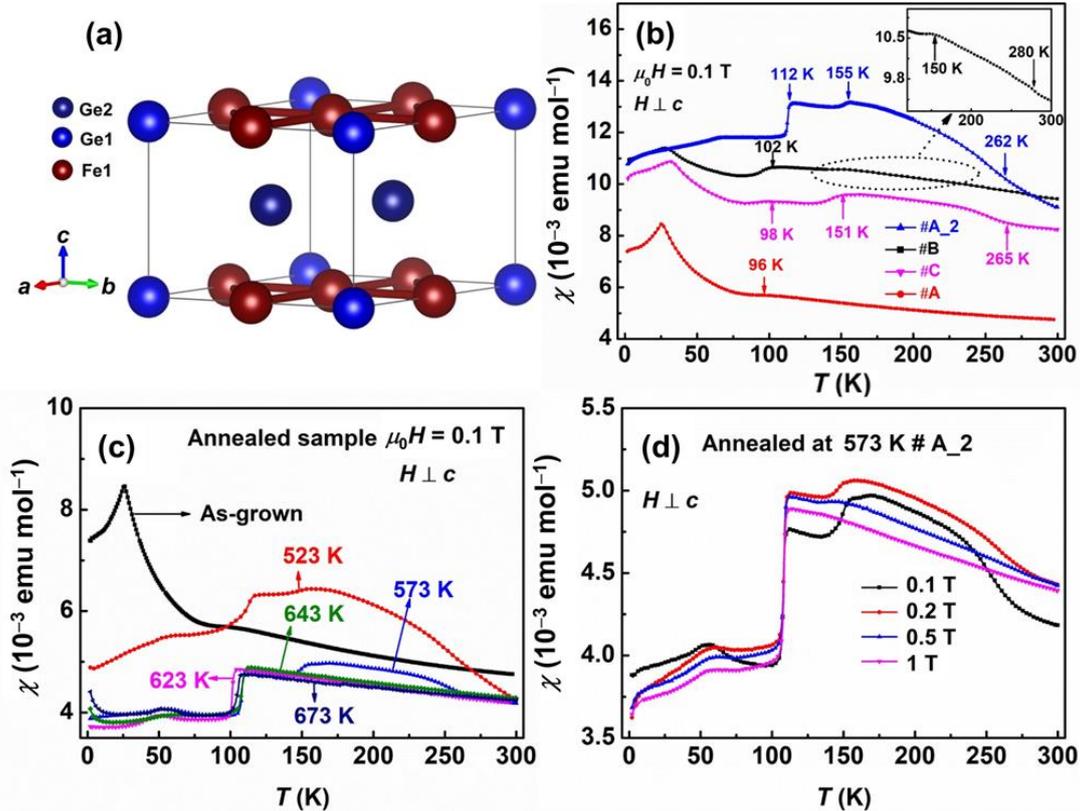

**Figure 1** (Color online) Crystal structures and magnetic susceptibility curves of FeGe. (a) Crystal structure of the B35-type FeGe as reported at room temperature. (b) Temperature-dependent magnetic susceptibility of FeGe crystals grown under different synthetic processes while crystal #A_2 (blue triangles) is an annealed crystal. The inset shows the close-up plot of magnetic susceptibility from 120 to 300 K for crystal #B. The temperatures of CDW order and other transitions from a secondary phase are marked. (c) Temperature-dependent magnetic susceptibility of FeGe samples annealed at different temperatures and the corresponding as-grown sample. (d) Temperature-dependent magnetic susceptibilities under different applied magnetic fields for the FeGe sample #A_2.

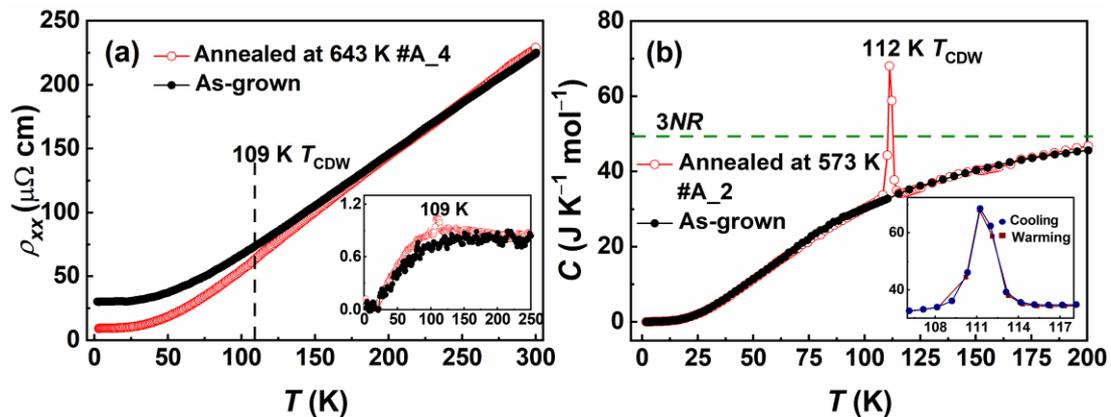

**Figure 2** (Color online) Resistivity and specific heat curves of FeGe. (a) Temperature-dependent in-plane electrical resistivity for as-grown (#A) and annealed (#A_4) samples. The inset shows the derivatives with respect to temperature around the CDW transition. (b) Temperature-dependent specific heat of as-grown (#A) and annealed (#A_2) samples. The green dashed line is the Dulong-Petit limit $3NR$ ($N$ is the number of the atoms in the formula, and $R$ is the gas constant) from phonon contribution. The inset shows the specific heat jump under the warming (brown) and cooling (blue) procedures.

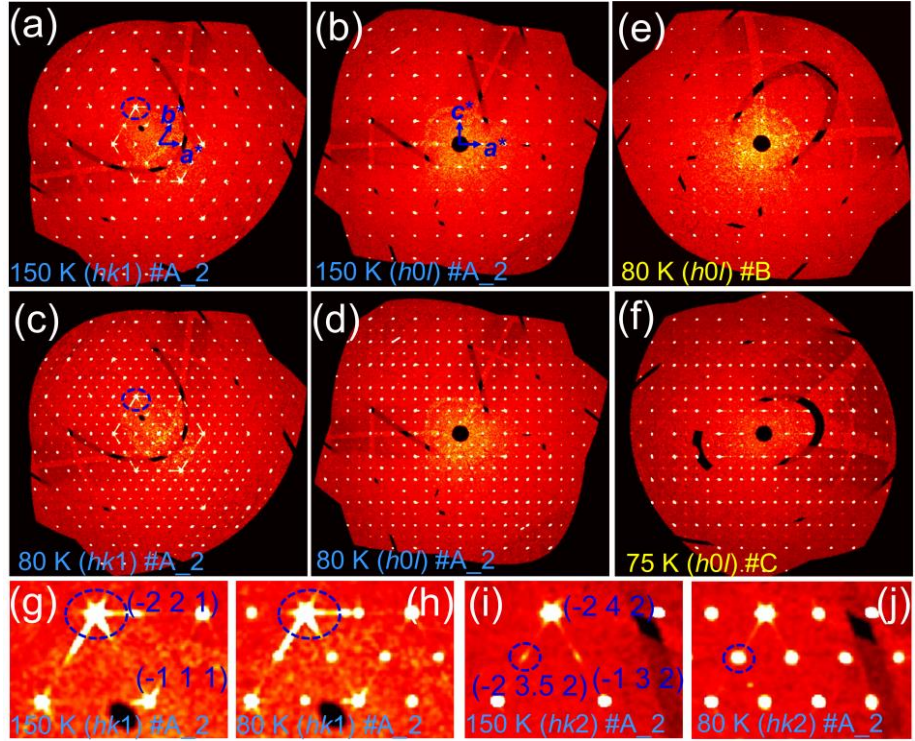

**Figure 3** (Color online) Reconstructed images for FeGe single crystals at different temperatures. (a-b) Reconstructed images of ($hk$1) and ($h0l$) planes in the reciprocal space at 150 K for the annealed FeGe sample #A_2, respectively. The reciprocal lattice unit vectors are marked. (c) (d) The corresponding images of the same planes below the phase transition at 80 K from the same sample in (a) and (b). (e) Reconstructed images of the ($h0l$) plane at 80 K for the as-grown FeGe sample #B. (f) Reconstructed images of ($h0l$) planes at 75 K for the as-grown FeGe sample #C. (g) (h) The zoomed areas of the diffuse scattering marked by the dashed circles in (a) and (c), respectively. (i) and (j) Reconstructed images of ($hk$2) planes in a small region for the sample #A_2 at 150 and 80 K, respectively. Some reflection indices are given. The black areas in the effective reciprocal spaces are due to the beam stop or the gaps between the chip modules in the detector.

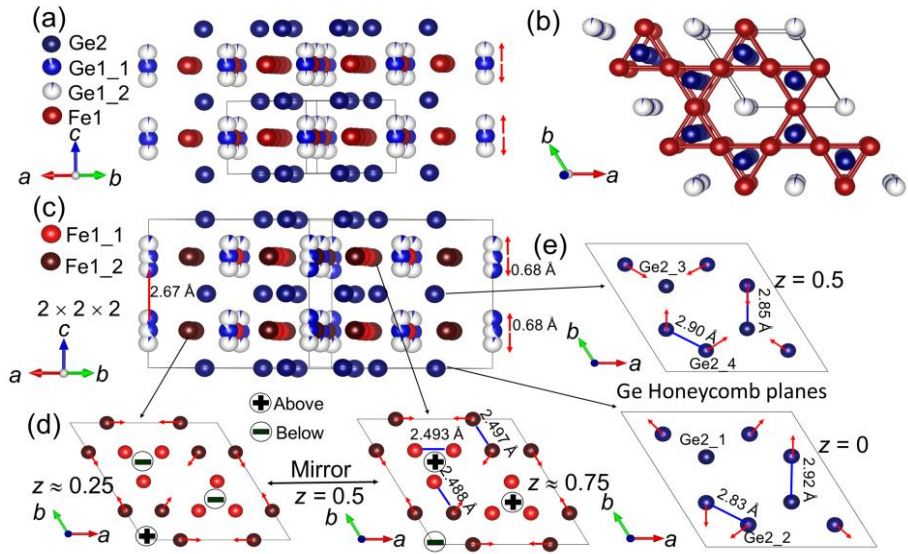

**Figure 4** (Color online) Crystal structures of annealed FeGe single crystals at 150 K and 80 K. (a) (b) Crystal structures of FeGe with disordered Ge1 sites at 150 K viewed along ab plane and $c$ axis, respectively. (c) Crystal structure of FeGe with a 2 × 2 × 2 supercell originating from CDW at 80 K. Disorder on Ge1 sites with different

occupancies is marked. The realization of Ge1 dimerization is illustrated by the arrows. The bond and displacement distances are noted. (d) The extremely weak distortion of Fe-Kagome patterns at $z \approx 0.25$ and $0.75$ from the $2 \times 2 \times 2$ supercell. They are symmetry-related by the mirror plane at $z = 0.5$. The in-plane and out-of-plane displacements of two different Fe sites are marked. The three different interatomic distances between two Fe atoms are labeled. (e) The weak distortion of Ge-honeycomb planes at $z = 0$ and $0.5$ from the $2 \times 2 \times 2$ supercell. The in-plane displacement of Ge atoms and four distinct interatomic distances between them are marked.

## 5.2 Tables

**Table 1** Crystal data and structure refinement for FeGe#A_2 at 80 K and 150 K

| Information related to the refinement | Results | |
|---|---|---|
| Sample label | FeGe#A_2 | |
| Formula weight | 128.4 g/mol | |
| Temperature | 80 K | 150 K |
| Wavelength | 0.5 Å | |
| Crystal system | hexagonal | |
| Space group | $P6/mmm$ | |
| Unit cell dimensions | $a = 9.9662(2)$ Å<br>$c = 8.0878(2)$ Å | $a = 4.9827(2)$ Å<br>$c = 4.0493(2)$ Å |
| Volume | 695.70(3) Å$^3$ | 87.064(7) Å$^3$ |
| Z | 24 | 3 |
| Density (calculated) | 7.3575 g/cm$^3$ | 7.3489 g/cm$^3$ |
| Absorption coefficient | 13.923 mm$^{-1}$ | 13.907 mm$^{-1}$ |
| $F(000)$ | 1392 | 174 |
| Crystal size | $0.059 \times 0.051 \times 0.012$ mm$^3$ | |
| $\theta$ range for data collection | 1.66 to 19.45° | 3.32 to 19.3° |
| Index ranges | $-13 \leq h \leq 13, -13 \leq k \leq 13, -10 \leq l \leq 10$ | $-6 \leq h \leq 6, -6 \leq k \leq 6, -5 \leq l \leq 5$ |
| Reflections collected | 19371 | 1310 |
| Independent reflections | 387 [$R_{int} = 0.0511$] | 60 [$R_{int} = 0.0301$] |
| Completeness to $\theta = 19.45°$ | 98% | 92% |
| Refinement method | $F$ | |
| Data / restraints / parameters | 387 / 0 / 41 | 60 / 0 / 10 |
| Goodness-of-fit | 1.63 | 0.93 |
| Final $R^*$ indices [$I>3\sigma(I)$] | $R_{obs} = 0.0219$, $wR_{obs} =$ | $R_{obs} = 0.0078$, $wR_{obs} =$ |
|  | 0.0298 | 0.0099 |
| $R$ indices [all data] | $R_{all}$ = 0.0254, $wR_{all}$ = 0.0304 | $R_{all}$ = 0.0078, $wR_{all}$ = 0.0099 |
| Extinction coefficient | 0.0650(90) | NA |
| Largest diff. peak and hole | 0.93 and –0.92 e·Å$^{-3}$ | 0.20 and –0.22 e·Å$^{-3}$ |

*$R = \Sigma||F_o|-|F_c||/\Sigma|F_o|$, $wR = \{\Sigma[w(|F_o|^2-|F_c|^2)^2]/\Sigma[w(|F_o|^4)]\}^{1/2}$ and $w=1/(\sigma^2(F)+0.0001F^2)$

**Acknowledgments**

Jin-Ke Bao acknowledges support from the National Natural Science Foundation of China (Grant No. 12204298). Shixun Cao would like to acknowledge the research grant from the National Natural Science Foundation of China (Grant No. 12074242), and the Science and Technology Commission of Shanghai Municipality (Grant No. 21JC1402600). Zhu-An Xu would like to acknowledge the National Natural Science Foundation of China (Grant No. 12174334). Yanpeng Qi would like to acknowledge the National Natural Science Foundation of China (Grant Nos. 52272265, U1932217, 11974246, 12004252). Xiaohui Yang would like to acknowledge the Zhejiang Provincial Natural Science Foundation of China under Grant No. LQ23A040009. The research at the University of Bayreuth has been funded by the Deutsche Forschungsgemeinschaft (DFG, German Research Foundation) – 406658237. We thank Martin Tolkiehn and Carsten Paulmann for their assistance at Beamline P24. We acknowledge DESY (Hamburg, Germany), a member of the Helmholtz Association HGF, for the provision of experimental facilities. Parts of this research were carried out at PETRA III, using beamline P24. Beamtime was allocated for proposal I-20220188.

**Conflict of Interest**   The authors declare that they have no conflict of interest.

# Supplementary Materials:
# Annealing-induced long-range charge density wave order in magnetic kagome FeGe: fluctuations and disordered structure

Chenfei Shi [1] [††], Yi Liu [2, 3] [††], Bishal Baran Maity [4], Qi Wang [5, 6], Surya Rohith Kotla [7], Sitaram Ramakrishnan [8], Claudio Eisele [7], Harshit Agarwal [7], Leila Noohinejad [9], Qian Tao [3], Baojuan Kang [1], Zhefeng Lou [10], Xiaohui Yang [11], Yanpeng Qi [5, 6, 12], Xiao Lin [10], Zhu-An Xu [3, 13], Arumugam Thamizhavel [4], Guang-Han Cao [3, 13], Sander van Smaalen [7] [*], Shixun Cao [1, 14] [*], and Jin-Ke Bao [1, 14, 15] [*]

[1] *Department of Physics, Materials Genome Institute, Institute for Quantum Science and Technology, Shanghai University, Shanghai 200444, China*
[2] *Department of Applied Physics, Zhejiang University of Technology, Hangzhou 310023, China*
[3] *School of Physics, Zhejiang Province Key Laboratory of Quantum Technology and Devices, Zhejiang University, Hangzhou 310027, China*
[4] *Department of Condensed Matter Physics and Materials Science, Tata Institute of Fundamental Research, Mumbai 400005, India*
[5] *School of Physical Science and Technology, ShanghaiTech University, Shanghai 201210, China*
[6] *ShanghaiTech Laboratory for Topological Physics, ShanghaiTech University, Shanghai 201210, China*
[7] *Laboratory of Crystallography, University of Bayreuth, Bayreuth 95447, Germany*
[8] *Department of Quantum Matter, AdSE, Hiroshima University, Higashi-Hiroshima 739-8530, Japan*
[9] *P24, PETRA III, Deutsches Elektronen-Synchrotron DESY, Hamburg 22607, Germany*
[10] *Key Laboratory for Quantum Materials of Zhejiang Province, School of Science, Westlake University, Hangzhou 310024, China*
[11] *Department of Physics, China Jiliang University, Hangzhou 310018, China*
[12] *Shanghai Key Laboratory of High-resolution Electron Microscopy, ShanghaiTech University, Shanghai 201210, China*
[13] *Collaborative Innovation Centre of Advanced Microstructures, Nanjing University, Nanjing 210093, China*
[14] *Shanghai Key Laboratory of High Temperature Superconductors, Shanghai University, Shanghai 200444, China*
[15] *School of Physics, Hangzhou Normal University, Hangzhou 311121, China*

[††] Chenfei Shi and Yi Liu contribute equally to this work. *Corresponding authors: Sander van Smaalen (smash@uni-bayreuth.de), Shixun Cao (sxcao@shu.edu.cn), and Jin-Ke Bao (baojk7139@gmail.com)

| Crystal information summary | |
|---|---|
| Sample label | Note on the growth information |
| #A | The as-grown sample batch in Zhejiang/Shanghai University |
| #B | The as-grown sample batch in Tata Institute of Fundamental Research |
| #C | The as-grown sample batch in ShanghaiTech University |
| #A_1 | Annealed at 523 K using the crystals from batch A |
| #A_2 | Annealed at 573 K using the crystals from batch A |
| #A_3 | Annealed at 623 K using the crystals from batch A |
| #A_4 | Annealed at 643 K using the crystals from batch A |
| #A_5 | Annealed at 673 K using the crystals from batch A |
| #A_6 | Annealed at 773 K using the crystals from batch A |

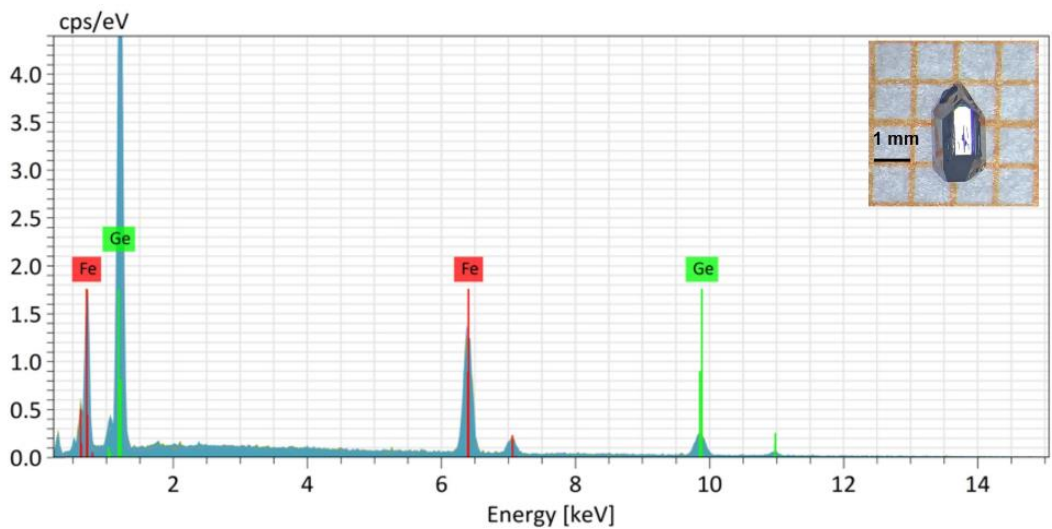

Figure S1 (Color online) EDS analysis of FeGe as-grown crystal #A showing that the atomic concentrations of Fe and Ge are 50 ± 2% and 50 ± 2%, respectively. Inset shows a photograph of FeGe crystal.

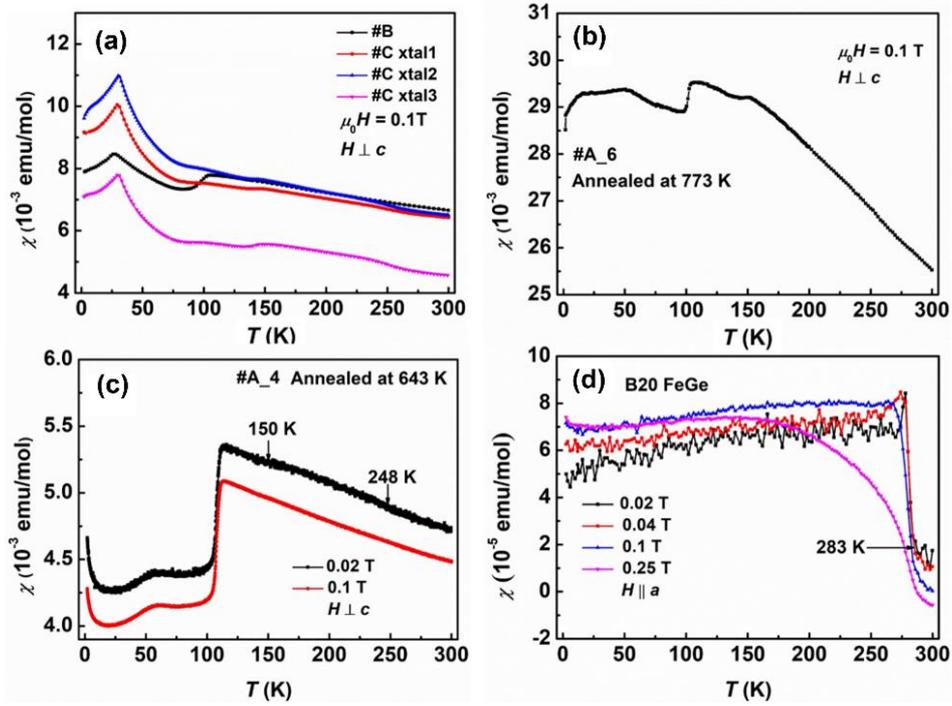

Figure S2 (Color online) Temperature-dependent magnetic susceptibilities for samples not shown in the Fig 1 of the main text. (a) crystals from batches #B and #C. (b) Batch #A_6 annealed at 773 K. The CDW transition becomes less pronounced for this annealing temperature. (c) Batch #A_4 annealed at 643 K under $\mu_0 H$ = 0.02 T and 0.1 T. Two weak anomalies at around 150 and 248 K can still be identified under a small field. (d) a cubic B20-type FeGe crystal under different magnetic fields. There is a ferromagnetic-like (helimagnetic) transition at around 283 K for the cubic FeGe phase.

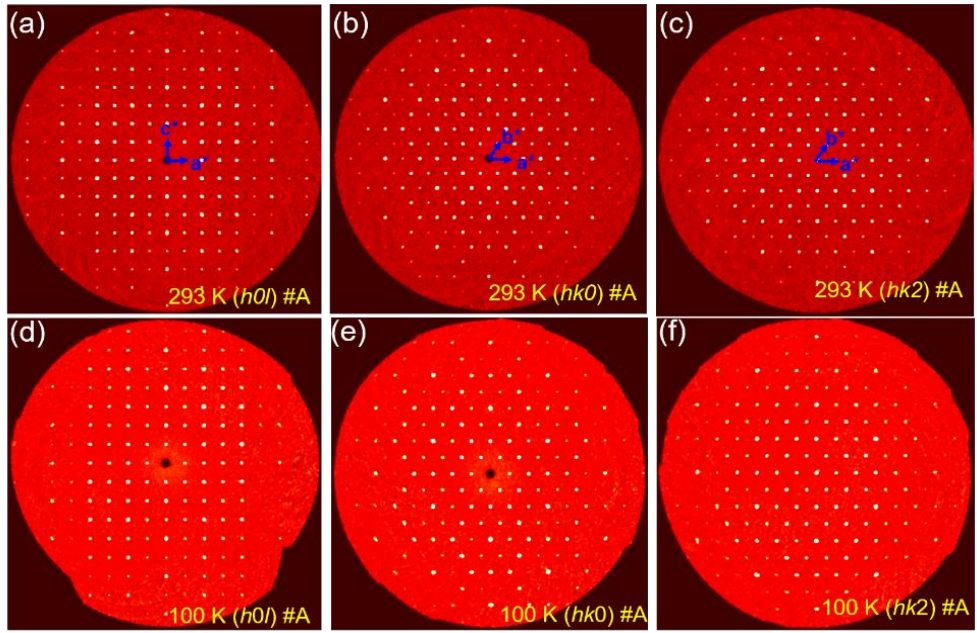

Figure S3 (Color online) (a)-(c) Reconstructed images of (*h0l*), (*hk*0) and (*hk*2) planes for reflections in the reciprocal space for the sample FeGe#A at 293 K and 100 K. The reciprocal lattice unit vectors are plotted.

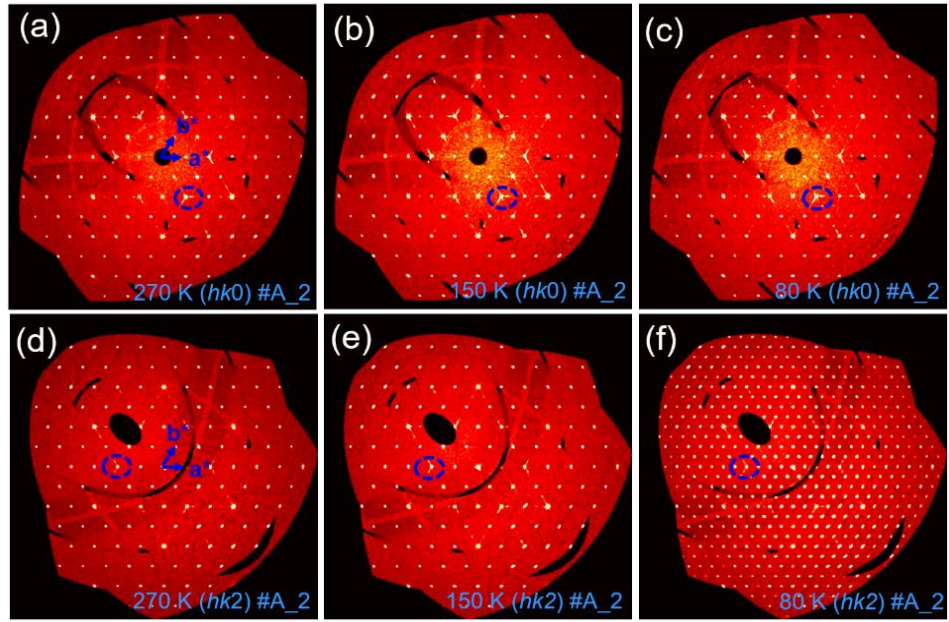

Figure S4 (Color online) (a)-(c) Reconstructed images of ($hk$0) planes for reflections in the reciprocal space for the sample FeGe#A_2 at 270, 150 and 80 K, respectively. Superlattice reflections from $2 \times 2$ supercell are weak in the ($hk$0) plane at 80 K. (d)-(f) Reconstructed images of ($hk$2) planes at 270, 150 and 80 K, respectively, for the same sample. Superlattice reflections are strong in the ($hk$2) plane at 80 K. The reciprocal lattice unit vectors are plotted in (a) and (d). Diffuse scattering features near some main reflections highlighted by blue dashed circles are persistent from 270 K across CDW ordering temperature to 80 K. The black areas in the effective reciprocal images are due to the beam stop or gaps between the chip modules in the detector.

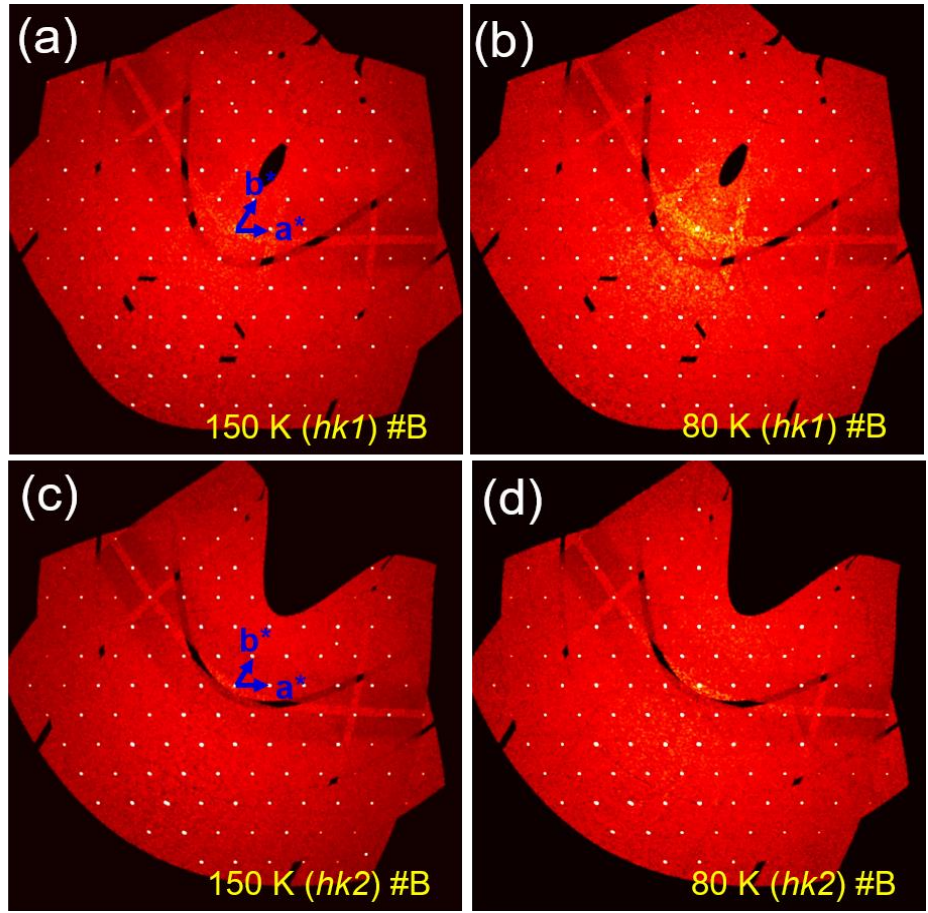

Figure S5 (Color online) (a) and (b) Reconstructed images of (*hk*1) plane for reflections in the reciprocal space for the sample FeGe#B at 150 and 80 K, respectively. (c) and (d) Reconstructed images of (*hk*2) plane for the same sample at 150 and 80 K, respectively. The reciprocal unit lattice vectors are plotted in (a) and (c). No obvious diffuse scattering feature are present in the sample FeGe#B. Almost no superlattice reflections are observed in those planes below the CDW ordering temperature at 80 K for the sample FeGe#B, either.

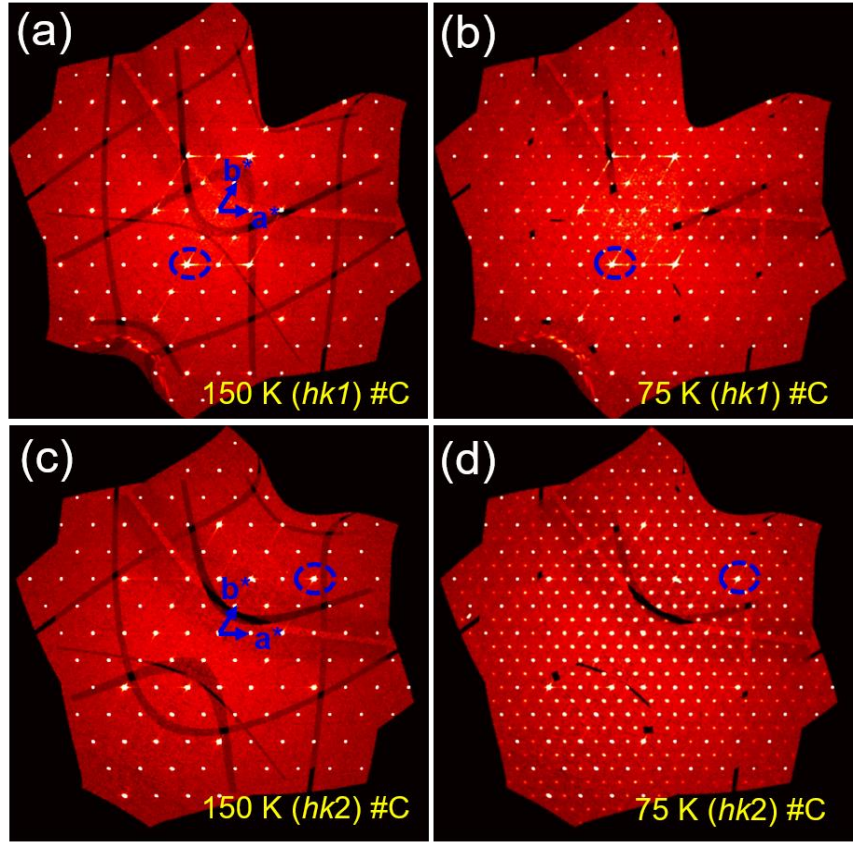

Figure S6 (Color online) (a) and (b) Reconstructed images of (*hk*1) plane for reflections in the reciprocal space for the sample FeGe#C at 150 and 75 K, respectively. (c) and (d) Reconstructed images of (*hk*2) plane for the same sample at 150 and 75 K, respectively. The reciprocal lattice unit vectors are plotted in (a) and (c). Superlattice reflections from 2 × 2 supercell are present at 75 K below the CDW ordering temperature. The diffuse scattering features highlighted by blue dashed circles persists from 150 to 75 K.

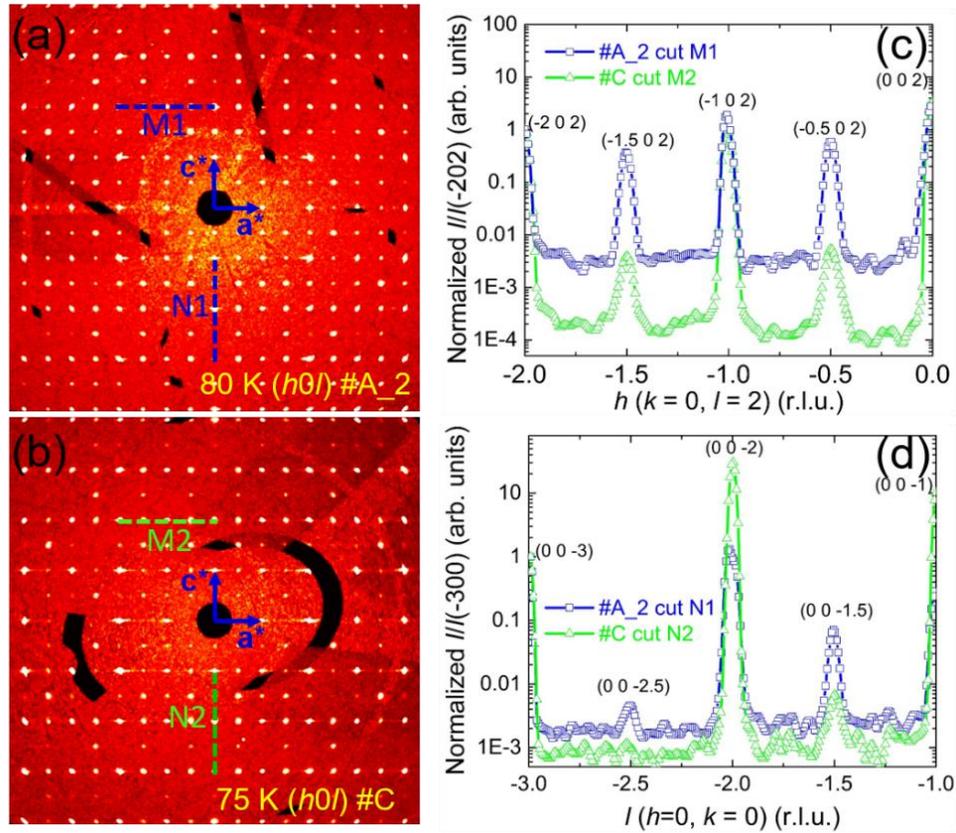

Figure S7 (Color online) (a) Reconstructed images of ($h0l$) plane for reflections in the reciprocal space at 80 K for the sample FeGe#A_2. (b) Reconstructed images of ($h0l$) at 75 K for the sample FeGe#C. The reciprocal lattice units are plotted. (c) Line cuts of M1 in (a) and M2 in (b) for the samples #A_2 and FeGe#C, respectively. (d) Line cuts of N1 in (a) and N2 in (b) for the samples FeGe#A_2 and FeGe#C, respectively. The indices for the reflections are labeled.

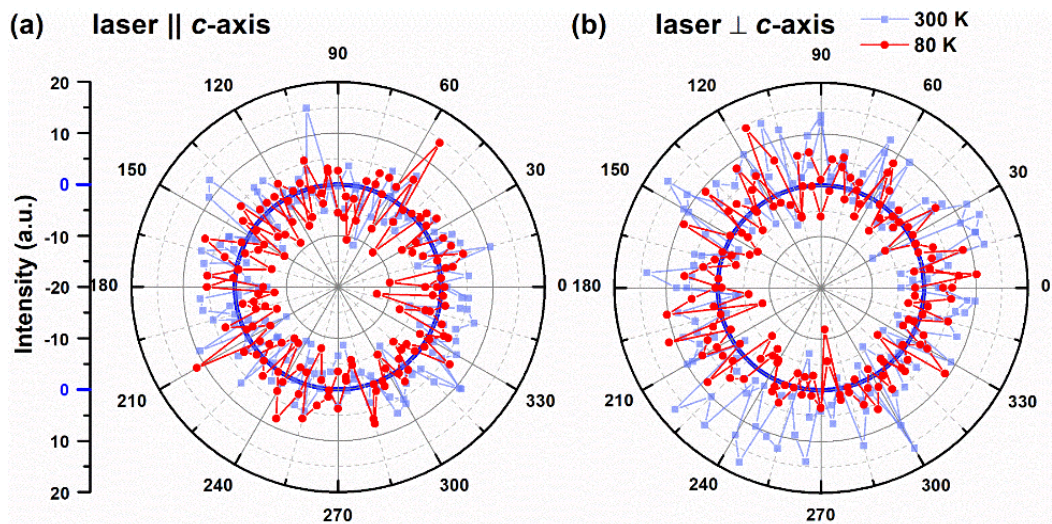

Figure S8 (Color online) The integrated second harmonic generation (SHG) signals as a function of the polarization angle $\theta$ for the annealed crystal #A_2 under 573 K measured above (300 K) and below (80 K) the CDW transition. The injected laser is (a) parallel to $c$-axis and (b) perpendicular to $c$-axis. No detectable SHG due to the inversion symmetry breaking is observed at 80 K.

Table S1. Crystal data and structure refinement for FeGe#**A** at 293 K.

| Sample label | FeGe#**A** | |
|---|---|---|
| Crystal model on Ge1 site | No disorder | With disorder |
| Formula weight | 127 g/mol | 128.4 g/mol |
| Temperature | 293 K | |
| Wavelength | 0.71073 Å | |
| Crystal system | hexagonal | |
| Space group | $P6/mmm$ | |
| Unit cell dimensions | $a = 4.9984(2)$ Å<br>$c = 4.0506(1)$ Å | |
| Volume | 87.642(5) Å$^3$ | |
| Z | 3 | |
| Density (calculated) | 7.2179 g/cm$^3$ | 7.3004 g/cm$^3$ |
| Absorption coefficient | 36.703 mm$^{-1}$ | 37.208 mm$^{-1}$ |
| $F(000)$ | 172 | 174 |
| Crystal size | 0.06 × 0.045 × 0.04 mm$^3$ | |
| $\theta$ range for data collection | 4.71 to 45.3° | |
| Index ranges | $-9 \leq h \leq 9, -9 \leq k \leq 9, -8 \leq l \leq 8$ | |
| Reflections collected | 9422 | |
| Independent reflections | 181 [$R_{int} = 0.0596$] | |
| Completeness to $\theta = 45.3°$ | 100% | |
| Refinement method | $F$ | |
| Data / restraints / parameters | 181 / 0 / 10 | 181 / 0 / 13 |
| Goodness-of-fit | 3.61 | 1.36 |
| Final $R$ indices [$I>3\sigma(I)$] | $R_{obs} = 0.0347$, $wR_{obs} = 0.0493$ | $R_{obs} = 0.0166$, $wR_{obs} = 0.0184$ |
| $R$ indices [all data] | $R_{all} = 0.0353$, $wR_{all} = 0.0493$ | $R_{all} = 0.0169$, $wR_{all} = 0.0184$ |
| Extinction coefficient | 0.0960(80) | 0.0990(30) |
| Largest diff. peak and hole | 9.08 and $-2.22$ e·Å$^{-3}$ | 1.34 and $-0.79$ e·Å$^{-3}$ |

$R = \Sigma||F_o|-|F_c|| / \Sigma|F_o|$, $wR = \{\Sigma[w(|F_o|^2 - |F_c|^2)^2] / \Sigma[w(|F_o|^4)]\}^{1/2}$ and $w=1/(\sigma^2(F)+0.0001F^2)$

Table S2. Atomic coordinates (×10$^4$) and equivalent isotropic displacement parameters (Å$^2$×10$^3$) for sample FeGe#A at 293 K with estimated standard deviations in parentheses.

| | | Label | x | y | z | Occupancy | $U_{eq}$* |
|---|---|---|---|---|---|---|---|
| FeGe#A | No disorder on Ge1 site | Ge(2) | 3333.33 | 6666.67 | 5000 | 1 | 8(1) |
| | | Ge(1) | 0 | 0 | 0 | 0.940(12) | 10(1) |
| | | Fe(1) | 5000 | 0 | 0 | 1 | 6(1) |
| | | Label | x | y | z | Occupancy | $U_{eq}$* |
| | With disorder on Ge1 site | Ge(2) | 3333.33 | 6666.67 | 5000 | 1 | 8(1) |
| | | Ge(1_1) | 0 | 0 | 0 | 0.880(3) | 9(1) |
| | | Fe(1) | 5000 | 0 | 0 | 1 | 6(1) |
| | | Ge(1_2) | 0 | 0 | 3030(30) | 0.0330(14) | 9 |
| | | Ge(1_3) | 0 | 0 | 1540(50) | 0.027 | 9 |

*$U_{eq}$ is defined as one third of the trace of the orthogonalized $U_{ij}$ tensor. Two independent Ge disordered sites are added in the final refinement. The following occupancy constraint are added: o[Ge(1_3)] = 0.5–0.5o[Ge(1_1)]–[Ge(1_2)].

Table S3. Crystal data and structure refinement for FeGe#**A** at 100 K.

| | |
|---|---|
| Empirical formula | FeGe#**A** |
| Formula weight | 128.4 g/mol |
| Temperature | 100 K |
| Wavelength | 0.71073 Å |
| Crystal system | hexagonal |
| Space group | *P*6/*mmm* |
| Unit cell dimensions | $a = 4.9843(11)$ Å, $\alpha = 90°$ <br> $c = 4.0471(9)$ Å, $\gamma = 120°$ |
| Volume | 87.07(3) Å$^3$ |
| Z | 3 |
| Density (calculated) | 7.3481 g/cm$^3$ |
| Absorption coefficient | 37.451 mm$^{-1}$ |
| *F*(000) | 175 |
| Crystal size | 0.06 × 0.045 × 0.04 mm$^3$ |
| $\theta$ range for data collection | 4.72 to 40.23° |
| Index ranges | $-9 \leq h \leq 9$, $-9 \leq k \leq 9$, $-7 \leq l \leq 7$ |
| Reflections collected | 6722 |
| Independent reflections | 142 [$R_{int}$ = 0.0713] |
| Completeness to $\theta = 40.23°$ | 100% |
| Refinement method | F |
| Data / restraints / parameters | 142 / 0 / 13 |
| Goodness-of-fit | 1.48 |
| Final *R* indices [$I>3\sigma(I)$] | $R_{obs} = 0.0176$, $wR_{obs} = 0.0204$ |
| *R* indices [all data] | $R_{all} = 0.0178$, $wR_{all} = 0.0204$ |
| Extinction coefficient | 0.0890(30) |
| Largest diff. peak and hole | 1.87 and $-1.81$ e·Å$^{-3}$ |

$R = \Sigma||F_o|-|F_c|| / \Sigma|F_o|$, $wR = \{\Sigma[w(|F_o|^2 - |F_c|^2)^2] / \Sigma[w(|F_o|^4)]\}^{1/2}$ and $w = 1/(\sigma^2(F)+0.0001F^2)$

Table S4. Atomic coordinates (×10$^4$) and equivalent isotropic displacement parameters (Å$^2$×10$^3$) for FeGe#**A** at 100 K with estimated standard deviations in parentheses.

| Label | x | y | z | Occupancy | $U_{eq}$* |
|---|---|---|---|---|---|
| Ge(2) | 3333.33 | 6666.67 | 5000 | 1 | 5(1) |
| Ge(1_1) | 0 | 0 | 0 | 0.802(4) | 4(1) |
| Fe(1) | 5000 | 0 | 0 | 1 | 4(1) |
| Ge(1_3) | 0 | 0 | 1620(30) | 0.0645(17) | 4(1) |
| Ge(1_2) | 0 | 0 | 3030(40) | 0.035(3) | 4(1) |

*$U_{eq}$ is defined as one third of the trace of the orthogonalized $U_{ij}$ tensor. Two independent Ge disordered sites are added in the final refinement. The following occupancy constraint are added: o[Ge(1_3)] = 0.5–0.5o[Ge(1_1)]–[Ge(1_2)].

Table S5. Crystal data and structure refinement for FeGe#**A_2** at 270 K and 150 K.

| Sample label | FeGe#**A_2** | |
|---|---|---|
| Formula weight | 128.4 g/mol | |
| Temperature | 270 K | 150 K |
| Wavelength | 0.5 Å | |
| Crystal system | hexagonal | |
| Space group | *P*6/*mmm* | |
| Unit cell dimensions | $a$ = 4.9929(3) Å<br>$c$ = 4.0525(3) Å | $a$ = 4.9827(2) Å<br>$c$ = 4.0493(2) Å |
| Volume | 87.490(10) Å$^3$ | 87.064(7) Å$^3$ |
| Z | 3 | |
| Density (calculated) | 7.3131 g/cm$^3$ | 7.3489 g/cm$^3$ |
| Absorption coefficient | 13.839 mm$^{-1}$ | 13.907 mm$^{-1}$ |
| $F$(000) | 174 | |
| Crystal size | 0.059 × 0.051 × 0.012 mm$^3$ | |
| $\theta$ range for data collection | 3.32 to 19.3° | |
| Index ranges | -6 ≤ $h$ ≤ 6, -6 ≤ $k$ ≤ 6, -5 ≤ $l$ ≤ 5 | |
| Reflections collected | 1337 | 1310 |
| Independent reflections | 62 [$R_{int}$ = 0.0278] | 60 [$R_{int}$ = 0.0301] |
| Completeness to $\theta$ = 19.3° | 94% | 92% |
| Refinement method | $F$ | |
| Data / restraints / parameters | 62 / 0 / 11 | 60 / 0 / 10 |
| Goodness-of-fit | 0.88 | 0.93 |
| Final $R$ indices [$I$>3σ($I$)] | $R_{obs}$ = 0.0074, $wR_{obs}$ = 0.0092 | $R_{obs}$ = 0.0078, $wR_{obs}$ = 0.0099 |
| $R$ indices [all data] | $R_{all}$ = 0.0074, $wR_{all}$ = 0.0092 | $R_{all}$ = 0.0078, $wR_{all}$ = 0.0099 |
| Extinction coefficient | 0.0610(40) | NA |
| Largest diff. peak and hole | 0.35 and –0.50 e·Å$^{-3}$ | 0.20 and –0.22 e·Å$^{-3}$ |

$R = \Sigma||F_o|-|F_c|| / \Sigma|F_o|$, $wR = \{\Sigma[w(|F_o|^2 - |F_c|^2)^2] / \Sigma[w(|F_o|^4)]\}^{1/2}$ and $w=1/(\sigma^2(F)+0.0001F^2)$

Table S6. Atomic coordinates (×10$^4$) and equivalent isotropic displacement parameters (Å$^2$×10$^3$) at 270 and 150 K for sample FeGe#**A_2** with estimated standard deviations in parentheses.

| | | Label | x | y | z | Occupancy | $U_{eq}$* |
|---|---|---|---|---|---|---|---|
| FeGe#**A_2** | 270 K | Ge(2) | 3333.33 | 6666.67 | 5000 | 1 | 6(1) |
| | | Ge(1_1) | 0 | 0 | 0 | 0.951(3) | 7(1) |
| | | Fe(1) | 5000 | 0 | 0 | 1 | 5(1) |
| | | Ge(1_2) | 0 | 0 | 1840(70) | 0.0245 | 7 |
| | 150 K | Label | x | y | z | Occupancy | $U_{eq}$* |
| | | Ge(2) | 3333.33 | 6666.67 | 5000 | 1 | 5(1) |
| | | Ge(1_1) | 0 | 0 | 0 | 0.938(3) | 5(1) |
| | | Fe(1) | 5000 | 0 | 0 | 1 | 4(1) |
| | | Ge(1_2) | 0 | 0 | 1790(60) | 0.0309 | 5 |

*$U_{eq}$ is defined as one third of the trace of the orthogonalized $U_{ij}$ tensor.
The following occupancy constraint is added: o[Ge(1_2)] = 0.5–0.5o[Ge(1_1)].

Table S7. The $R_{obs}$ factors of all the reflections and specific groups of superlattice reflections under different space groups for the sample FeGe#A_2 at 80 K. The crystal lattice parameters are chosen as the 2×2×2 cell as the high temperature phase.

| Space group | Centro-symmetric | $R_{obs}$(all the reflections) | $R_{obs}$(h = 2n+1) | $R_{obs}$(l = 2n+1) | $R_{obs}$(satellite reflections) | Note |
|---|---|---|---|---|---|---|
| P6/mmm | Yes | 2.19% | 3.27% | 4.20% | 3.47% | |
| P-62m | No | 2.17% | 3.39% | 4.94% | 3.71% | Non-positive definite ADP |
| P6mm | No | 2.06% | 3.43% | 4.28% | 3.6% | Non-positive definite ADP |
| P-6m2 | No | 2.56% | 3.9% | 4.79% | 4.08% | Non-positive definite ADP |
| P622 | No | 2.17% | 3.21% | 6.61% | 3.9% | |

"obs" is defined reflections as $I>3\sigma(I)$. $R_{obs}(h = 2n+1)$ and $R_{obs}(l = 2n+1)$ refer to all the reflections with an index of $h = 2n+1$ and $l = 2n+1$, respectively. They belong to superlattice reflections which appear below $T_{CDW}$. ADP is short for atomic displacement parameter.

Table S8. Atomic coordinates (×10$^4$) and equivalent isotropic displacement parameters (Å$^2$×10$^3$) for FeGe#A_2 at 80 K with estimated standard deviations in parentheses.

| Label | x | y | z | Occupancy | $U_{eq}$* |
|---|---|---|---|---|---|
| Ge(2_1) | 3333.33 | 6666.67 | 0 | 1 | 4(1) |
| Ge(2_2) | 1641(1) | 3282 | 0 | 1 | 5(1) |
| Ge(2_3) | 3333.33 | 6666.67 | 5000 | 1 | 4(1) |
| Ge(2_4) | 1684(1) | 3367 | 5000 | 1 | 5(1) |
| Ge(1_1_2) | 5000 | 0 | 2488(1) | 0.887(6) | 5(1) |
| Ge(1_1_1) | 0 | 0 | 2510(5) | 0.388(3) | 3(1) |
| Fe(1_1) | 2499(1) | 4999 | 2484(1) | 1 | 4(1) |
| Fe(1_2) | 2505(1) | 0 | 2516(1) | 1 | 4(1) |
| Ge(1_2_1) | 0 | 0 | 3350(3) | 0.532(3) | 3 |
| Ge(1_2_4) | 5000 | 0 | 1660(20) | 0.075 | 5 |
| Ge(1_2_2) | 0 | 0 | 1660(20) | 0.080 | 3 |
| Ge(1_2_3) | 5000 | 0 | 3310(50) | 0.039(7) | 5 |

*$U_{eq}$ is defined as one third of the trace of the orthogonalized $U_{ij}$ tensor.
The following occupancy constraints are added: o[Ge(1_2_4)] = 1–o[Ge(1_1_2)]–o[Ge(1_2_3)], o[Ge(1_2_2)] = 1–o[Ge(1_1_1)]–o[Ge(1_2_1)].

Table S9. Crystal data and structure refinement for FeGe#**B** at 150 and 80 K.

| Sample label | FeGe#**B** | |
|---|---|---|
| Formula weight | 128.4 g/mol | |
| Temperature | 150 K | 80 K |
| Wavelength | 0.5 Å | |
| Crystal system | hexagonal | |
| Space group | *P*6/*mmm* | |
| Unit cell dimensions | $a$ = 4.98860(10) Å $c$ = 4.05360(10) Å | $a$ = 4.98750(10) Å $c$ = 4.0502(2) Å |
| Volume | 87.363(3) Å$^3$ | 87.252(5) Å$^3$ |
| Z | 3 | |
| Density (calculated) | 7.3237 g/cm$^3$ | 7.3331 g/cm$^3$ |
| Absorption coefficient | 13.859 mm$^{-1}$ | 13.877 mm$^{-1}$ |
| $F$(000) | 174 | |
| Crystal size | 0.068 × 0.064 × 0.045 mm$^3$ | |
| $\theta$ range for data collection | 3.32 to 20.21° | |
| Index ranges | -6 ≤ $h$ ≤ 6, -6 ≤ $k$ ≤ 6, -5 ≤ $l$ ≤ 5 | |
| Reflections collected | 1049 | 1981 (including two runs) |
| Independent reflections | 68 [$R_{int}$ = 0.0514] | 68 [$R_{int}$ = 0.0587] |
| Completeness to $\theta$ = 20.21° | 100% | 99% |
| Refinement method | $F$ | |
| Data / restraints / parameters | 68 / 0 / 11 | 68 / 0 / 11 |
| Goodness-of-fit | 1.54 | 3.03 |
| Final $R$ indices [$I$>3σ($I$)] | $R_{obs}$ = 0.0211, $wR_{obs}$ = 0.0227 | $R_{obs}$ = 0.0359, $wR_{obs}$ = 0.0401 |
| $R$ indices [all data] | $R_{all}$ = 0.0223, $wR_{all}$ = 0.0229 | $R_{all}$ = 0.0359, $wR_{all}$ = 0.0401 |
| Extinction coefficient | 0.1160(80) | 0.0900(140) |
| Largest diff. peak and hole | 1.37 and –1.45 e·Å$^{-3}$ | 2.54 and –2.45 e·Å$^{-3}$ |

$R = \Sigma||F_o|-|F_c|| / \Sigma|F_o|$, $wR = \{\Sigma[w(|F_o|^2 - |F_c|^2)^2] / \Sigma[w(|F_o|^4)]\}^{1/2}$ and $w=1/(\sigma^2(F)+0.0001F^2)$

Table S10. Atomic coordinates (×10⁴) and equivalent isotropic displacement parameters (Å²×10³) for FeGe#**B** at 150 K with estimated standard deviations in parentheses.

| | | Label | $x$ | $y$ | $z$ | Occupancy | $U_{eq}$* |
|---|---|---|---|---|---|---|---|
| FeGe#**B** | 150 K | Ge(2) | 3333.33 | 6666.67 | 5000 | 1 | 5(1) |
| | | Ge(1_1) | 0 | 0 | 0 | 0.910(6) | 8(1) |
| | | Fe(1) | 5000 | 0 | 0 | 1 | 4(1) |
| | | Ge(1_2) | 0 | 0 | 2350(80) | 0.045 | 8 |
| | | Label | $x$ | $y$ | $z$ | Occupancy | $U_{eq}$* |
| | 80 K | Ge(2) | 3333.33 | 6666.67 | 5000 | 1 | 5(1) |
| | | Ge(1_1) | 0 | 0 | 0 | 0.832(10) | 8(1) |
| | | Fe(1) | 5000 | 0 | 0 | 1 | 4(1) |
| | | Ge(1_2) | 0 | 0 | 2030(80) | 0.084 | 8 |

*$U_{eq}$ is defined as one third of the trace of the orthogonalized $U_{ij}$ tensor.
The following occupancy constraint is added: o[Ge(1_2)] = 0.5–0.5o[Ge(1_1)].

Table S11. Crystal data and structure refinement for FeGe#**C** at 150 K.

| Sample label | FeGe#**C** | |
|---|---|---|
| Formula weight | 128.4 g/mol | |
| Temperature | 150 K | 75 K |
| Wavelength | 0.5 Å | |
| Crystal system | hexagonal | |
| Space group | $P6/mmm$ | $P6/mmm$ (average structure) |
| Unit cell dimensions | $a$ = 4.98640(10) Å<br>$c$ = 4.05220(10) Å | $a$ = 4.9859(2) Å<br>$c$ = 4.0496(2) Å |
| Volume | 87.256(3) Å$^3$ | 87.183(7) Å$^3$ |
| Z | 3 | |
| Density (calculated) | 7.3327 g/cm$^3$ | 7.3389 g/cm$^3$ |
| Absorption coefficient | 13.876 mm$^{-1}$ | 13.888 mm$^{-1}$ |
| $F(000)$ | 174 | 174 |
| Crystal size | 0.058 × 0.034 × 0.030 mm$^3$ | |
| $\theta$ range for data collection | 3.32 to 19.3° | |
| Index ranges | $-6 \leq h \leq 6, -6 \leq k \leq 6, -5 \leq l \leq 5$ | |
| Reflections collected | 1442 | 1835 (including two runs) |
| Independent reflections | 65 [$R_{int}$ = 0.0303] | 61 [$R_{int}$ = 0.0269] |
| Completeness to $\theta$ = 19.3° | 100% | 92% |
| Refinement method | $F$ | |
| Data / restraints / parameters | 65 / 0 / 11 | 61 / 0 / 11 |
| Goodness-of-fit | 1.40 | 1.95 |
| Final $R$ indices [$I>3\sigma(I)$] | $R_{obs}$ = 0.0120, $wR_{obs}$ = 0.0149 | $R_{obs}$ = 0.0143, $wR_{obs}$ = 0.0195 |
| $R$ indices [all data] | $R_{all}$ = 0.0120, $wR_{all}$ = 0.0149 | $R_{all}$ = 0.0143, $wR_{all}$ = 0.0195 |
| Extinction coefficient | 0.0360(40) | 0.0300(110) |
| Largest diff. peak and hole | 1.10 and –0.44 e·Å$^{-3}$ | 1.73 and –0.92 e·Å$^{-3}$ |

$R = \Sigma ||F_o|-|F_c|| / \Sigma |F_o|$, $wR = \{\Sigma[w(|F_o|^2 - |F_c|^2)^2] / \Sigma[w(|F_o|^4)]\}^{1/2}$ and $w=1/(\sigma^2(F)+0.0001F^2)$

Table S12. Atomic coordinates (×10⁴) and equivalent isotropic displacement parameters (Å²×10³) for FeGe#**C** at 150 and 75 K (average structure) with estimated standard deviations in parentheses.

| | | Label | $x$ | $y$ | $z$ | Occupancy | $U_{eq}$* |
|---|---|---|---|---|---|---|---|
| FeGe#**C** | 150 K | Ge(2) | 3333.33 | 6666.67 | 5000 | 1 | 4(1) |
| | | Ge(1_1) | 0 | 0 | 0 | 0.911(3) | 6(1) |
| | | Fe(1) | 5000 | 0 | 0 | 1 | 3(1) |
| | | Ge(1_2) | 0 | 0 | 2430(40) | 0.0444 | 6 |
| | 75 K | Label | $x$ | $y$ | $z$ | Occupancy | $U_{eq}$* |
| | | Ge(2) | 3333.33 | 6666.67 | 5000 | 1 | 4(1) |
| | | Ge(1_1) | 0 | 0 | 0 | 0.833(4) | 6(1) |
| | | Fe(1) | 5000 | 0 | 0 | 1 | 2(1) |
| | | Ge(1_2) | 0 | 0 | 2060(30) | 0.084 | 6 |

*$U_{eq}$ is defined as one third of the trace of the orthogonalized $U_{ij}$ tensor.
The following occupancy constraint is added: o[Ge(1_2)] = 0.5–0.5o[Ge(1_1)].